\documentclass[12pt]{article}
\usepackage{amsfonts}
\usepackage{graphicx}
\usepackage{epstopdf}
\usepackage{epsfig}
\usepackage{amssymb}
\usepackage{setspace}
\usepackage{caption}
\usepackage{amsfonts}
\usepackage{color}
\usepackage{amsmath}
\usepackage{float}
\usepackage{comment}
\newcommand {\be}{\begin{equation}}
\newcommand {\ee}{\end{equation}}
 \newcommand {\bea}{\begin{array}}
 
 \newcommand {\eea}{\end{array}}

 \newcommand {\RN}{Reissner–Nordstrom~}

\textwidth=6.5in \hoffset=-.75in \voffset=-1in \numberwithin{equation}{section}
\numberwithin{figure}{section}

\begin{document}

\begin{titlepage}
	\vspace{1cm}
	\begin{center}
		{\Large \bf {Magnetized Kerr-Taub-NUT spacetime and Kerr/CFT correspondence}}\\
	\end{center}
	\vspace{2cm}
	\begin{center}
		\renewcommand{\thefootnote}{\fnsymbol{footnote}}
		Haryanto M. Siahaan{\footnote{haryanto.siahaan@unpar.ac.id}}\\
		Center for Theoretical Physics,\\
		Department of Physics, Parahyangan Catholic University,\\
		Jalan Ciumbuleuit 94, Bandung 40141, Indonesia
		\renewcommand{\thefootnote}{\arabic{footnote}}
	\end{center}
	
	\begin{abstract}
		
We present a new solution in Einstein-Maxwell theory which can be considered as the magnetized version of Kerr-Taub-NUT solution. Some properties of the spacetime are discussed. We also compute the entropy of extremal black hole in the spacetime by using the Kerr/CFT correspondence approach. 
		
	\end{abstract}
\end{titlepage}\onecolumn
\bigskip

\section{Introduction}
\label{sec.intro}

Kerr spacetime is an exact solution to the vacuum Einstein equations describing the spacetime outside a rotating mass \cite{Wald:1984rg.book,Griffiths:2009dfa}. Adding the NUT parameter to this solution yields the Kerr-Taub-NUT (KTN) spacetime which also belongs to the vacuum Einstein system. Interestingly, the presence of NUT parameter in the solution leads to a regular Kretschmann scalar at the origin, in contrast to the generic Kerr black hole solution that possesses a ring singularity inside the horizon. However, despite the non existence of ring singularity hiding inside the horizon of KTN black hole, KTN spacetime suffers the so-called conical singularity \cite{Griffiths:2009dfa} which is related to the non $2\pi$ periodicity of angular coordinate $\phi$. In addition to this conic singularity, another peculiar feature of a spacetime equipped with NUT parameter is the closed timelike curve (CTC) \cite{Griffiths:2009dfa} where a spacelike coordinate with a periodicity can become timelike. Nonetheless, in spite of these obscure properties of KTN spacetime, numerous studies have been performed to explore its aspects \cite{Yang:2020iat,Duztas:2017lxk,Frolov:2017bdq,Jefremov:2016dpi,Cebeci:2015fie,Pradhan:2013hqa,Cebeci:2012mb}. Even a spacetime solution with NUT parameter can also exist in theories beyond Einstein-Maxwell \cite{Cisterna:2021xxq,Siahaan:2020bga,Siahaan:2019kbw}.

Adding electromagnetic field into the spacetime of Einstein-Maxwell theory yield an electrovacuum system. One of the popular exact solutions to the Einstein-Maxwell equations is the Kerr-Newman spacetime. This solution describes spacetime outside a rotating charged and massive object. Moreover, this solution can contain a black hole provided that the rotating collapsing object can maintain some amount of electric charge during its gravitational contraction \cite{Wald:1984rg.book}. Another way to add electromagnetic fields in the spacetime is by magnetizing a solution following the Ersnt formalism \cite{Ernst}. Basically, this Ersnt magnetization is a type of Harrison transformation acting on some known seed solutions in Einstein-Maxwell theory. The review on Ersnt magnetization can be found in literature \cite{Aliev:1989wz,Aliev:1989sw,Aliev:1989wx}, and one can view the Melvin magnetic universe \cite{Melvin:1963qx} can be obtained by magnetizing the Minkowski spacetime. 

Particularly, applying the Ernst magnetization to the Kerr spacetime gives us the magnetized Kerr solution \cite{Ernst} which is interpreted as the spacetime outside a Kerr black hole immersed in a strong magnetic field. Unlike the generic Kerr solution which has the asymptotic flatness, the magnetized version is no longer flat at asymptotic due to the presence of homogeneous external magnetic field filling in the spacetime. Indeed, this nature of magnetized Kerr solution is found quite unrealistic to be associated to any known astrophysical system, but one can still consider it as some approximate way to model a black hole immersed in a strong magnetic field coming from the accretion disc \cite{Aliev:1989sw}. Therefore, it can be understood why this solution is still worth further investigations, for example in its relation to the Meisner effect \cite{Bicak:2015lxa} and the Kerr/CFT holography \cite{Siahaan:2015xia,Astorino:2015naa}.  Some other aspects of magnetized black holes can be found in literature \cite{Aliev:1989wx,Brito:2014nja,Kolos:2015iva,Tursunov:2014loa,Astorino:2016hls,Orekhov:2016bpc,Astorino:2015lca,Booth:2015nwa,Gibbons:2013yq,Gibbons:2013dna,Rogatko:2016knj}

The Kerr/CFT correspondence has been used as a tool to compute the entropy of extremal Kerr black hole using some two dimensional conformal field theory (CFT$_2$) point of views \cite{Guica:2008mu,Hartman:2008pb,Compere:2012jk}. The $SL\left(2,{\mathbb R}\right) \times U\left(1\right)$ isometry of the near horizon geometry of the extremal black hole is the window where the CFT$_2$ method can apply. The obtained diffeomorphisms subject to some boundary conditions can be brought to be a set of Virasoro generators of a CFT$_2$ whose commutator incorporates a specific central charge. In turn, plugging the obtained central charge and Frolov-Thorne temperature in Cardy formula, the entropy for extremal black hole can be reproduced. This Kerr/CFT holography has been applied to many type of rotating or charged black holes in the Einstein-Maxwell theory \cite{Compere:2012jk} and beyond \cite{Ghezelbash:2009gf,Azeyanagi:2008kb}. 

Despite the real world application of the NUT parameter or external magnetic field in the spacetime obeying the Einstien-Maxwell equations is still vague, the discussions of spacetime exhibiting these properties have extended our understanding of some aspects in gravitational theories \cite{Hawking:1976jb,Hawking:1998ct,Bena:2005ni,Luna:2015paa,Page:1979aj,Bena:2009ev}. In particular in how we define the conserved quantities in the spacetime such as mass and angular momentum, and also the geodesics of test particles in the spacetime \cite{Jefremov:2016dpi}. Also in a recent work \cite{Ciambelli:2020qny}, the authors show that the Misner string contribution to the entropy of Taub-NUT-AdS can be renormalized by introducing the Gauss-Bonnet term. Therefore, one may wonder what the properties of spacetime with both NUT parameter and external magnetic field. This is exactly the motivation in \cite{Siahaan:2021ags} where the author introduced the magnetized version of the Taub-NUT solution, which was extended to the charged case namely the \RN-Taub-NUT spacetime \cite{Siahaan:2021ypk}. In this paper, still in the same spirit, we would like apply the magnetization scheme to the case of KTN spacetime to obtain the magnetized KTN (MKTN) solution which is to the best of our knowledge does not exist in literature. 

The organization of this paper is as follows. In the next section, we construct the MKTN solution by employing the Ernst magnetization to the KTN metric as the seed solution. Some aspects of the obtained spacetime such as the surface area, squared of Riemann tensor at equator, and the Hawking temperature are studied in section \ref{sec.aspects}. In section \ref{sec.microentropy}, the near horizon geometry of the black hole is obtained, followed by the microscopic entropy calculation according to Kerr/CFT correspondence. Finally we give conclusions and discussions. A brief review of Ernst magnetization is presented in the appendix \ref{app.ErnstMag}. We consider the natural units $c={\hbar} = k_B = G_4 = 1$.

\section{Magnetized Kerr-Taub-NUT spacetime}
\label{sec.MKerrTN}

The MKTN spacetime presented in this section is a result of Ernst magnetization to the KTN line element
\be \label{metric.KTN}
d{s^2} =  - {{{\Delta _r}{{\left( {dt - \left( {a{\Delta _x} - 2lx} \right)d\phi } \right)}^2}} \over \Xi} + {\rho ^2}\left( {{{d{r^2}} \over {{\Delta _r}}} + {{d{x^2}} \over {{\Delta _x}}}} \right) + {{{\Delta _x}} \over {\Xi}}{\left( {adt - \left( {{r^2} + {a^2} + {l^2}} \right)d\phi } \right)^2}
\ee 
as the seed. Above we have $\Xi = {{{r^2} + {\left( {l + ax} \right)^2}}}$, ${\Delta _r} = {r^2} - 2mr + {a^2} - {l^2}$, and ${\Delta _x} = 1 - {x^2}$. Note that the KTN (\ref{metric.KTN}) solution above solves the vacuum Einstein equations, so there is no electromagnetic field associated to the system. To apply the Ersnt magnetization, first we need to rewrite the last metric into the Lewis-Papapetrou-Weyl form (\ref{LPWmetric1}), where the corresponding functions can be found as the followings,
\be 
f = {{{\Delta _x}{{\left( {{r^2} + {l^2} + {a^2}} \right)}^2} - {\Delta _r}{{\left( {a{\Delta _x} - 2lx} \right)}^2}} \over {{\Xi}}}\,,
\ee 
\be 
\omega  = {{{\Delta _r}\left( {a{\Delta _x} - 2lx} \right) - a{\Delta _x}\left( {{r^2} + {l^2} + {a^2}} \right)} \over {{\Delta _r}{{\left( {a{\Delta _x} - 2lx} \right)}^2} - {\Delta _x}{{\left( {{r^2} + {l^2} + {a^2}} \right)}^2}}}\,,
\ee 
\be 
e^{2\gamma} = {{\Delta _x}{{\left( {{r^2} + {l^2} + {a^2}} \right)}^2} - {\Delta _r}{{\left( {a{\Delta _x} - 2lx} \right)}^2}}\,,
\ee 
and $\rho^2 = \Delta_x \Delta_r$. In the absence of electromagnetic field, the existing Ernst potential is only $\cal E$, where the real and imaginary components are given by
\[ 
{{\mathop{\rm Re}\nolimits} {\cal E}} =\frac{1}{\Xi} \left\{ \Delta_x r^4 + \left({a}^{2}+4lax-6{l}^{2}{x}^{2}-{a}^{2}{x}^{4}+2{l}^{2}-4a{x}^{3}l
\right) r^2 + 2m \left( 2lx+a{x}^{2}-a \right) ^{2} r\right.
\]
\be 
\left. - \left( l+ax \right)  \left( {a}^{3}{x}^{3}-{a}^{3}x-a{x}^{3}{l}^{2}-
3{l}^{3}{x}^{2}+3{a}^{2}{x}^{2}l+5{l}^{2}ax-3{a}^{2}l-{l}^{3}\right)  \right\}\,,
\ee 
and
\[ 
 {\mathop{\rm Im}\nolimits} {\cal E} = \frac{1}{a^2 \Xi} \left\{ 2{a}^{2}l \left( {x}^{2}+1 \right) r^3 -2m \left( {a}^{3}{x}^{3}-3{a}^{3}x-2{a}^{2}l-{l}^{3}+3{a}^{2}{x}^{2}l \right) r^2 + 2{a}^{2}l \left( 3{a}^{2}{x}^{2} \right. \right.
\]
\be 
\left. \left. -2a{x}^{3}l+{l}^{2}+6lax-{a}^{2}-3{l}^{2}{x}^{2} \right) r + 2m \left( {a}^{2}+{l}^{2} \right)  \left( l+ax \right)  \left( {a}^{2}{x}^{2}+{a}^{2}+lax+{l}^{2} \right)  \right\}\,,
\ee
respectively. Following (\ref{LambdaDEF}), the $\Lambda$ function associated to this Ernst potential is given by
\be 
\Lambda  = {{{{\tilde \Lambda }_R} + i{{\tilde \Lambda }_I}} \over {{a^2}\left( {l + ax + ir} \right)}}\,,
\ee 
where
\[{{\tilde \Lambda }_R} = -{a}^{2}{b}^{2} \left( l+3l{x}^{2}-ax\Delta_x \right)  r^2 + 2{b}^{2}m \left( {a}^{3}{x}^{3}-3{a}^{3}x-{l}^{3}-2{a}^{2}l+3{a}^{2}l{x}^{2} \right) r \]
\be 
-a^2 \left\{ \left( ab^2 x^3 + 3b^2 x^2 l \right) \left(a^2 - l^2\right)-ax \left(b^{2}{a}^{2}+1-5 {b}^{2}{l}^{2}\right) -l \left( 3{b}^{2}{a}^{2}+1+{b}^{2}{l}^{2} \right) \right\}\,,
\ee 
and
\[
{{\tilde \Lambda }_I} = b^2 a^2 \Delta_x r^3 + \left\{{b}^{2} \left( 3\,{l}^{2}+{a}^{2} \right) {x}^{2}-4a{b}^{2}lx-{b}^{2}{a}^{2}-3{b}^{2}{l}^{2}-1 \right\} r
\] 
\be 
2{b}^{2}m \left( {l}^{2}+{a}^{2} \right)  \left( {a}^{2}{x}^{2}+{a}^{2}+lax+{l}^{2} \right) \,.
\ee 

Then the transformed Ernst potentials, which give us a set of magnetized solution for the metric and vector potential, can be found using (\ref{magnetization}). The resulting magnetized metric can be expressed as
\be\label{mag-metric} 
{\rm{d}}s^2  = \frac{1}{{f'}}\left\{ { - \rho ^2 {\rm{d}}t^2  + e^{2\gamma } \left( {\frac{{{\rm{d}}r^2 }}{{\Delta _r }} + \frac{{{\rm{d}}x^2 }}{{\Delta _x }}} \right)} \right\} + f'\left( {{\rm{d}}\phi  - \omega '{\rm{d}}t} \right)^2  \,,
\ee 
where $f' = {\left| \Lambda  \right|^2}f$, and
\be 
\omega ' = {{\sum\limits_{j = 0}^6 {{c_j}{x^j}} } \over {\sum\limits_{k = 0}^4 {{d_k}{x^k}} }}\,,
\ee 
with the corresponding $c_j$'s are given by
\[
c_6 = 2 \left( {l}^{2}{a}^{2}+3mr{a}^{2}-2{a}^{2}{m}^{2}+3{l}^{2}mr-
{r}^{3}m-3{r}^{2}{l}^{2}-{l}^{4}-2{l}^{2}{m}^{2} \right) {a}^{5}{b}^{4} \Delta_r\,,
\]
\[
c_5 = -2 \left( 3{a}^{4}-6{l}^{2}{a}^{2}+8{a}^{2}{m}^{2}-6{r}^{2}{a}^{2}-8mr{a}^{2}-8{l}^{2}mr+3{l}^{4}-{r}^{4}+6{r}^{2}{l}^{2}+
4{r}^{3}m+8{l}^{2}{m}^{2} \right) {a}^{4}{b}^{4}l \Delta_r\,,
\]
\[
c_4 = -2 \left( 12{l}^{4}{m}^{2}-15{l}^{4}{a}^{2}-12mr{l}^{4}+16{a}^{4}{l}^{2}-15mr{l}^{2}{a}^{2}-21{r}^{2}{l}^{2}{a}^{2}+14{m}^{2}{a}^{2}{l}^{2}\right.
\]
\[
\left. -7m{r}^{3}{a}^{2}-3mr{a}^{4}+4{m}^{2}{a}^{4}\right) {a}^{3}{b}^{4}\Delta_r\,,
\]
\[
c_3 = 4 \left( {l}^{4}{a}^{2}+6mr{l}^{4}-4{l}^{4}{m}^{2}-20{a}^{4}{l}^{2}+8mr{l}^{2}{a}^{2}+6{r}^{2}{l}^{2}{a}^{2}-8{m}^{2}{a}^{2}{l}^{2}+2m{r}^{3}{l}^{2}+{a}^{6}+2mr{a}^{4}\right. 
\]
\[
\left. +{a}^{2}{r}^{4}+6m{r}^{3}{a}^{2}-6{r}^{2}{a}^{4}-8{m}^{2}{a}^{4} \right) {a}^{2}{b}^{4}l \Delta_r\,,
\]
\[
c_2 = -3 a^4 m b^4 r^5  -6{a}^{4}{b}^{4} \left( 3{l}^{2}-2{m}^{2} \right) r^4 -2{a}^{2}{b}^{4}m \left( 5{a}^{4}+4{l}^{4}-19{l}^{2}{a}^{2} \right) r^3 -6{a}^{2}{b}^{4} \left({a}^{4}{l}^{2} -4{l}^{4}{m}^{2}-4{m}^{2}{a}^{4} \right.
\]
\[
\left. -6{m}^{2}{a}^{2}{l}^{2} \right) r^2 -m \left\{m^2 4{b}^{4} \left( {l}^{2}+3{a}^{2} \right)  \left( -{l}^{4}-3{l}^{2}{a}^{2}+{a}^{4} \right) {a}^{2} \left(7{a}^{6}{b}^{4} -{a}^{2}-69{a}^{2}{b}^{4}{l}^{4}+34{a}^{4}{b}^{4}{l}^{2}-24{b}^{4}{l}^{6} \right) \right\} r\,,
\]
\[
c_1 = 2 l \Delta_r \left\{-3r^4 a^4 b^4 -4{a}^{2}{b}^{4}m \left( {l}^{2}+2{a}^{2} \right) r^3 +6a^4 b^4 \left(a^2 - l^2\right) r^2 -12{a}^{2}{b}^{4}m \left( {a}^{2}+{l}^{2} \right) ^{2} r \right.
\]
\[
\left. +{a}^{8}{b}^{4}+{b}^{4}{a}^{4}{l}^{4}+18{a}^{6}{b}^{4}{l}^{2}+8{a}^
{6}{b}^{4}{m}^{2}-12{a}^{4}{b}^{4}{m}^{2}{l}^{2}-{a}^{4}-4{b}^{4}{
	m}^{2}{l}^{6}-16{a}^{2}{b}^{4}{m}^{2}{l}^{4}
 \right\}\,,
\]
\[
c_0 = 2a\left\{ -3{a}^{4}m{b}^{4} {r}^{5} -2{a}^{2}{b}^{4}m \left( 2{l}^{4}+7{l}^{2}{a}^{2}+{a}^{4} \right) r^3 + {a}^{2}{b}^{4} \left(3{a}^{4}{l}^{2} -8{m}^{2}{a}^{2}{l}^{2}-4{l}^{4}{m}^{2}-6{m}^{2}{a}^{4} \right) r^2 
 \right.
\]
\[
+m \left({a}^{8}{b}^{4} -16{a}^{4}{b}^{4}{m}^{2}{l}^{2}-{a}^{4}-8{a}^{6}{b}^{4}{l}^{2}-27{b}^{4}{a}^{4}{l}^{4}-12{b}^{4}{l}^{6}{a}^{2}-4{b}^{4}{m}^{2}{l}^{6}-16{a}^{2}{b}^{4}{m}^{2}{l}^{4} \right) r
\]
\[
\left. -4{b}^{4}{m}^{2}{l}^{8}+ 4{a}^{2}{b}^{4} \left( {a}^{2}-3{m}^{2} \right) {l}^{6}+ \left( 13{a}^{6}{b}^{4}-4{a}^{4}{b}^{4}{m
}^{2} \right) {l}^{4}+ \left( 3{a}^{8}{b}^{4}+10{a}^{6}{b}^{4}{m}^
{2}-{a}^{4} \right) {l}^{2}+2{a}^{8}{b}^{4}{m}^{2}
 \right\}\,,
\]
and the corresponding $d_k$'s are
\[
d_4 = a^6 \Delta_r\,,
\]
\[
d_3 = 4l a^5 \Delta_r\,,
\]
\[
d_2 = - \left( 8{l}^{2}mr+{a}^{4}-4mr{a}^{2}-{r}^{4}-8{l}^{2}{a}^{2}+3{l}^{4}-6{r}^{2}{l}^{2} \right) {a}^{4}\,,
\]
\[
d_1 = -4la^5 \Delta_r\,,
\]
\[
d_0 = - \left( 2mr{a}^{2}+{r}^{2}{a}^{2}+3{l}^{2}{a}^{2}+{r}^{4}+{l}^{4}
+2{r}^{2}{l}^{2} \right) {a}^{4}\,.
\]
Note that the magnetized metric (\ref{mag-metric}) possesses the Killing vectors $\zeta^\mu_{\left(t\right)}\partial_\mu = \partial_t$ and $\zeta^\mu_{\left(\phi\right)}\partial_\mu = \partial_\phi$ just like the seed solution (\ref{metric.KTN}). 

Unlike the seed solution which belongs to the vacuum Einstein, the magnetized version is now contain electromagnetic field where the corresponding Ernst potential is given by
\be \label{mag.Phi}
\Phi ' = {{{{\cal F}_R} + i{{\cal F}_I}} \over {{{\cal G}_R} + i{{\cal G}_I}}}\,,
\ee 
where
\[
{\cal F}_R = b \left\{ {a}^{2} \left( l-ax+a{x}^{3}+3 l{x}^{2} \right) r^2 -2\,m \left( -{l}^{3}+{a}^{3}{x}^{3}-3{a}^{3}x-2{a}^{2}l+3{a}^{2}l{x}^{2} \right) r  \right.
\]
\be 
\left. + {a}^{2} \left( {a}^{3}{x}^{3} -{l}^{3}-a{l}^{2}{x}^{3}+5a{l}^{2}x-3{l}^{3}{x}^{2}
-{a}^{3}x+3{a}^{2}l{x}^{2}-3{a}^{2}l \right) \right\}\,,
\ee 
\be 
{\cal F}_I = -b \left\{ a^2 \Delta_x r^2 -{a}^{2} \left( 3 {l}^{2}{x}^{2}-{a}^{2}\Delta_x-3{l}^{2}-4 lax \right)r + m \left( {a}^{2}+{l}^{2} \right)  \left( {a}^{2}{x}^{2}+{a}^{2}+lax+{l}^{2} \right) 
\right\}\,,
\ee 
\[
{\cal G}_R = -{a}^{2}{b}^{2} \left(l-ax\Delta_x +3l{x}^{2} \right) r^2 + 2{b}^{2}m \left( 3
{a}^{2}l{x}^{2} -{l}^{3}+{a}^{3}{x}^{3}-3{a}^{3}x-2{a}^{2}l \right) r -{a}^{2} \left( {b}^{2}{a}^{3}{x}^{3} \right. 
\]
\be 
\left. -{l}^{3}{b}^{2}-{b}^{2}{a}^{3}x-3
{b}^{2}{a}^{2}l-3{x}^{2}{l}^{3}{b}^{2}-l-{b}^{2}a{l}^{2}{x}^{3}+5
{b}^{2}a{l}^{2}x+3{b}^{2}{a}^{2}{x}^{2}l-ax \right)\,,
\ee 
\[
{\cal G}_I = b^2 a^2 \Delta_x r^2 -{a}^{2} \left( 3{x}^{2}{b}^{2}{l}^{2}-4a{b}^{2}lx-3{l}^{2}{b}^{2}-{b}^{2}{a}^{2}-1+{b}^{2}{a}^{2}{x}^{2} \right) r 
\]
\be 
+ 2{b}^{2}m \left( {a}^{2}+{l}^{2} \right)\left({a}^{2}{x}^{2}+{a}^{2}+lax+{l}^{2} \right) \,.
\ee 
From this potential, the components of vector field $A_\mu$ can be obtained using eq. (\ref{Ernst.potential.EM}), (\ref{drAt}), and (\ref{dxAt}) accordingly. However, the full expression of $A_\phi$ and ${\tilde A}_\phi$ are tedious, and we provide them in appendix \ref{app.vec}. Furthermore, getting the full expression for $A_t$ is troublesome since we need to integrate $A_\phi$ and ${\tilde A}_\phi$. Therefore, since in this present work we do not need the full mathematical expression for $A_\mu$, we will omit to present the complete result for $A_t$. In using some symbolic manipulation programs we can show the obtained solution solves the Einstein-Maxwell equations by using the field strength tensor $F_{\mu\nu}$ constructed from the $A_\phi$ solution in (\ref{Ap}) and $\partial_r A_t$ together with $\partial_x A_t$ obeying (\ref{drAt}) and (\ref{dxAt}), respectively.

\section{Some aspects of the black hole}\label{sec.aspects}

The spacetime (\ref{mag-metric}) introduced in this paper can contain a black hole with radius $r_+ = m + \sqrt{m^2-a^2+l^2}$, exactly the same as the radius of KTN horizon. The total area of the black hole then can be computed using the standard textbook formula
\be \label{area}
{\cal A} = \int\limits_0^{2\pi } {d\phi \int\limits_{ - 1}^1 {dx\sqrt {g_{\phi \phi } g_{xx} } } }  = 4\pi \left(r_ + ^2 + a^2+ l^2\right) \,,
\ee
which is equal to the area of KTN black hole \cite{Pradhan:2013hqa}. Following the relation where the entropy of a black hole is a quarter of its area \cite{Pradhan:2013hqa}, the entropy of a MKTN black hole reads
\be \label{entropy.gen}
S = \frac{\cal A}{4} = \pi \left(r_ + ^2 + a^2 + l^2\right) \,.
\ee 

\begin{figure}
	\centering
	\includegraphics[scale=0.4]{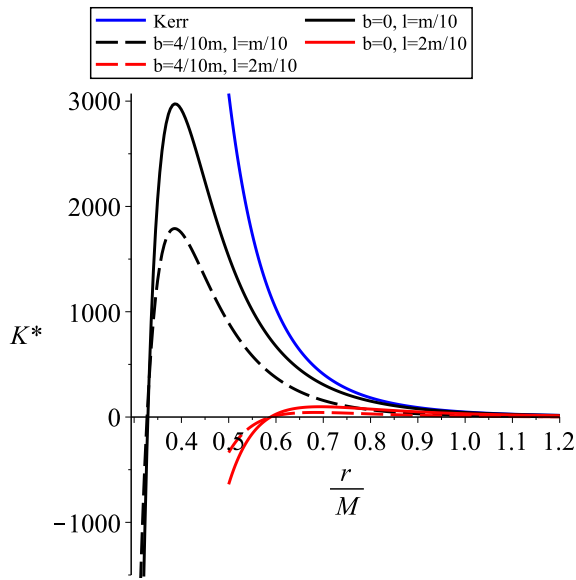}
	\caption{Plots of dimensionless squared Riemann tensor $K^* = m^{-4} R_{\alpha\beta\mu\nu} R^{\alpha\beta\mu\nu}$ at $x=0$ for the spacetime (\ref{mag-metric}) where we have considered the numerical value $a=0.2~ m$.}\label{fig.K}
\end{figure}

To grasp more aspects of this new spacetime, let us here examine the squared of corresponding Riemann curvature tensor. Surely expressing this quantity in its full complicated and lengthy expression, even on equator, is unnecessary. Then we would rather to present some of its numerical evaluations as presented in fig. \ref{fig.K}. The typical curvature in the spacetime with NUT parameter is obvious in the red and black lines, where they intersect the vertical axis of $r=0$ at some finite values of $K^* = m^{-4} R_{\alpha\beta\mu\nu} R^{\alpha\beta\mu\nu}$. We can compare to the blue line which represents the curvature in generic Kerr spacetime, where there exist singularity at the origin. This confirms our earlier statement that rotating spacetimes with NUT parameter, including MKTN, does not possess the true ring singularity as the null NUT parameter counterpart exhibits.  

Now let us turn to a semiclassical aspect where we would like to compute the Hawking temperature associated to a black hole in MKTN spacetime. Here we adopt the complex path method proposed in \cite{Srinivasan:1998ty} which has been applied to a Taub-NUT black hole in \cite{Kerner:2006vu}. We start with a general form of metric
\be \label{metric.Kerner}
 ds^2  =  - \tilde f\left( {r,x} \right)dt^2  + \frac{{dr^2 }}{{\tilde g\left( {r,x} \right)}} + \tilde C\left( {r,x} \right)h_{ij} \left( {r,x} \right)d\tilde x^i d\tilde x^j\,,
\ee 
where $\tilde x^i  = \left[ {x,\tilde \phi } \right]$, and ${\tilde \phi} = \phi - \omega' t$. Obviously the metric (\ref{mag-metric}) can be rewritten in the form (\ref{metric.Kerner}) above, hence the complex path approach for Hawking radiation presented in \cite{Kerner:2006vu} can apply. Now we consider the geodesic $dx = d{\tilde \phi}=0$, for a fixed $x=0$, then the $\left(t-r\right)$ sector of metric above which matters. In such consideration and using the Hamilton-Jacobi ansatz for the scalar field $\Phi = \exp\left[-iS\left(t,r\right)\right]$, the reading of massless Klein-Gordon equation $\nabla_\mu \nabla^\mu \Phi$ can be written in the form
\be \label{eqS}
\left( {\frac{{\partial S}}{{\partial r}}} \right)^2  = \frac{1}{{\tilde f\left( r \right)\tilde g\left( r \right)}}\left( {\frac{{\partial S}}{{\partial t}}} \right)^2 \,.
\ee  
As the spacetime being stationary, one can look for a solution
\be \label{Sansatz}
S\left( {t,r} \right) = Et + \tilde S\left( r \right)\,.
\ee 
Then the solution to eq. (\ref{eqS}) can be written as 
\be \label{sol.S}
S\left( {t,r} \right) = E\left( {t \pm \int\limits_0^r {\frac{{dr}}{{\sqrt {\tilde f\left( r \right)\tilde g\left( r \right)} }}} } \right)\,.
\ee 
Plugging the last result into our Hamilton-Jacobi ansatz yields the ingoing and outgoing modes
\be 
\Phi _{{\rm{in}}}  = \exp \left[ { - iE\left( {t + \int\limits_0^r {\frac{{dr}}{{\sqrt {\tilde f\left( r \right)\tilde g\left( r \right)} }}} } \right)} \right]\,,
\ee 
and
\be 
\Phi _{{\rm{out}}}  = \exp \left[ { - iE\left( {t - \int\limits_0^r {\frac{{dr}}{{\sqrt {\tilde f\left( r \right)\tilde g\left( r \right)} }}} } \right)} \right]\,,
\ee 
respectively. 

Obviously, the probability of ingoing particle is unity, i.e. $
P_{{\mathop{\rm in}\nolimits} }  = \left| {\Phi _{{\rm{in}}} } \right|^2  = 1$. Then by using the detailed balance principle
\be 
\frac{{P_{{\rm{out}}} }}{{P_{{\rm{in}}} }} = \exp \left[ { - \frac{E}{{T_H }}} \right]\,,
\ee
one can write the Hawking temperature
\be 
T_H  = \frac{1}{4}\left( {{\mathop{\rm Im}\nolimits} \int\limits_0^r {\frac{{dr}}{{\sqrt {\tilde f\left( r \right)\tilde g\left( r \right)} }}} } \right)^{ - 1} \,.
\ee 
Plugging the corresponding functions from the MKTN metric (\ref{mag-metric}) into the last equation gives us
\be 
T_H  = \frac{1}{4}\left( {{\mathop{\rm Im}\nolimits} \int\limits_0^r {\frac{{\sqrt {\left( {r^2  + a^2  + l^2 } \right)^2  - a^2 \Delta _r } }}{{\left( {r - r_ +  } \right)\left( {r - r_ -  } \right)}}} } \right)^{ - 1}\,, 
\ee 
which yields the Hawking temperature for a MKTN black hole can be written as
\be \label{TH}
T_H  = \frac{{r_ +   - r_-  }}{{4\pi \left( {r_ + ^2 + a^2 + l^2 } \right)}}\,.
\ee 
As we have expected, this Hawking temperature is the same to that of KTN case \cite{Pradhan:2013hqa}. As an alternative, the Hawking temperature above can also be achieved by using the straightforward calculation of surface gravity
\be 
\kappa^2  =  - \frac{1}{2}\left( {\nabla _\mu  \xi _\nu  } \right)\left( {\nabla ^\mu  \xi ^\nu  } \right)\,.
\ee 
The result is
\be 
\kappa = \frac{{r_ +   - r_-  }}{{2 \left( {r_ + ^2 + a^2 + l^2 } \right)}}\,,
\ee 
and after using the relation $T_H = \tfrac{\kappa}{2\pi}$, we can confirm the Hawking temperature (\ref{TH}) above.

\section{Microscopic entropy for extremal black holes}\label{sec.microentropy}

In this section, we investigate the Kerr/CFT correspondence for a MKTN black hole. We would like to extent the previous works \cite{Siahaan:2015xia,Astorino:2015naa} where the authors show how to compute the entropy of an extremal magnetized Kerr black hole using the Kerr/CFT holography method. The initial step is to find the near horizon geometry of the extremal black hole which possesses the $SL\left(2,{\mathbb R}\right)\times U\left(1\right)$ isometry. This near horizon metric together with the corresponding near horizon vector field solution solve the Eintein-Maxwell equations. Luckily, the general formulation of asymptotic symmetry group (ASG) method in Kerr/CFT correspondence for the Einstein-Maxwell theory has been discussed in \cite{Hartman:2008pb}, and had been extended to a class of gravitational theory in \cite{Compere:2012jk}. Hence, in this section we just need to obtain a near horizon geometry which falls into the category considered in \cite{Hartman:2008pb}, and employ the general formula for central charge given there.

The near horizon geometry of a MKTN black hole can be achieved by performing the following coordinate transformation 
\be 
t \to \frac{{r_0 t}}{\lambda }~~,~~r \to r_e  + \lambda r_0 r~~,~~\phi  \to \phi  + \Omega _J^{ext} \frac{{r_0 }}{\lambda }t\,.
\ee 
In the equations above we have $\Omega_J = \omega'$, and $\Omega _J^{ext}$ is the corresponding quantity evaluated at extremality, i.e. $m^2+l^2 = a^2$. By applying this near horizon transformation to the metric (\ref{mag-metric}), one can obtain the near horizon line element
\be \label{nhmetric}
{\rm{d}}s^2  = \Gamma \left( x \right)\left\{  - r^2 {\rm{d}}t^2  + \frac{{{\rm{d}}r^2 }}{{r^2 }} + \alpha \left( x \right)\frac{{{\rm{d}}x^2 }}{{\Delta _x }} \right\} + \gamma \left( x \right)\left( {{\rm{d}}\phi  + kr{\rm{d}}t} \right)^2  \,,
\ee 
after setting $r_0 = \sqrt{2} a$. In equation ( \ref{nhmetric}) we have $\alpha\left(x\right) =1$,
\[
\Gamma \left( x \right) = a^{-3}\left\{  \left[4 \left( 4{a}^{8}-{l}^{8}-3{l}^{4}{a}^{4}-5{l}^{6}{a}^{2}+9
{l}^{2}{a}^{6} \right) b^4 -{a}^{4} \left( 8{a}^{2}{b}^{2}-1 \right) \right] a x^2 + \left[ 4 \left( {l}^{2}{a}^{6} -{l}^{8}+8{a}^{8} \right. \right. \right.
\]
\be 
\left. \left. \left. -7{l}^{4}{a}^{4}-5{l}^{6}{a}^{2} \right) b^4  +  a^4 \right] 2lx + 4a \left( 4{a}^{8}-{l}^{8}-3{l}^{4}{a}^{4}-5{l}^{6}{a}^{2}+9{l}^{2}{a}^{6} \right)b^4 + {a}^{5} \left( 8{a}^{2}{b}^{2}+1 \right) \right\}\,,
\ee 
\be 
\gamma \left( x \right) = \frac{4 a^4 \Delta_x}{\Gamma\left(x\right)}\,,
\ee 
and
\be\label{k} 
k = - \frac{\sqrt{a^2-l^2}}{a^5} \left\{ 4{a}^{6}{b}^{4} \left( 4{a}^{2}+3{l}^{2} \right)- {a}^{4} \left(1-  28{b}^{4}{l}^{4} \right)+4{l}^{6}{b}^{4} \left( 5{a}^{2}+{l}^{2} \right)\right\}\,.
\ee

To complete the metric solution above in solving the Einstein-Maxwell equations, the accompanying vector field is given by
\be \label{nhA}
A_\mu  {\rm{d}}x^\mu   = L\left( x \right)\left( {{\rm{d}}\phi  + kr{\rm{d}}t} \right)\,,
\ee 
where
\be 
L\left( x \right) = \frac{L_1 x + L_2}{{\Gamma \left( x \right)}{\left\{4 {b}^{4} \left(9{a}^{6}{l}^{2}-3{a}^{4}{l}^{4}-5{a}^{2}{l}^{6}+4{a}^{8}-{l}^{8} \right) +{a}^{4} \left(1- 8{a}^{2}{b}^{2} \right) \right\}}}\,,
\ee
where
\be 
L_1 = a \left(36{a}^{6}{b}^{4}{l}^{2} -12{a}^{4}{b}^{4}{l}^{4}-8{a}^{6}{b}^{2}-20{a}^{2}{l}^{6}{b}^{4}+{a}^{4}-4{l}^{8}{b}^{4}+16{a}^{8}{b}^{4} \right)\,, 
\ee 
and 
\be 
L_2 = l \left( 4{a}^{6}{b}^{4}{l}^{2}+32{a}^{8}{b}^{4}-20{a}^{2}{l}^{6}{b}^{4}+{a}^{4}-28{a}^{4}{b}^{4}{l}^{4}-4{l}^{8}{b}^{4} \right) \,.
\ee 
As we aimed, this near horizon geometry possesses the $SL\left(2,{\mathbb R}\right)\times U\left(1\right)$ isometry which is vital for Kerr/CFT correspondence to apply. The $SL\left(2,{\mathbb R}\right)$ symmetry is generated by the Killing vectors
\be 
K _ -   = \partial _t \,,
\ee 
\be 
K _0  = t\partial _t  - r\partial _r \,,
\ee 
\be 
K _ +   = \left( {\frac{1}{{2r^2 }} + \frac{{t^2 }}{2}} \right)\partial _t  - tr\partial _r  - \frac{k}{r}\partial _\phi  \,,
\ee 
which obey the commutation relations $\left[ {K _0 ,K _ \pm  } \right] =  \pm K _ \pm   $ and $\left[ {K_ -  ,K_ +  } \right] = K_0 $, while the $U\left(1\right)$ symmetry is generated by $ \partial_\phi$. Here we learn that the presence of NUT parameter does not break the $SL\left(2,{\mathbb R}\right)\times U\left(1\right)$ symmetry of the near horizon extremal magnetized Kerr black hole studied in \cite{Siahaan:2015xia,Astorino:2015naa}. 

Moreover, the form of the near horizon metric (\ref{nhmetric}) and vector fields (\ref{nhA}) match the general forms of near horizon Einstein-Maxwell fields discussed in \cite{Hartman:2008pb,Kunduri:2007vf}. Hence, we can apply the ASG method constructed in \cite{Hartman:2008pb,Compere:2012jk} to our present case. Note that we also need to impose the following boundary condition for the spacetime metric
\be 
h_{\mu \nu }  \sim \left( {\begin{array}{*{20}c}
		{{\cal O}\left( {r^2 } \right)} & {{\cal O}\left( {r^{ 1} } \right)} & {{\cal O}\left( {r^{ - 1} } \right)} & {\cal O}\left( r^{-2} \right)  \\
		{} & {{\cal O}\left(1\right)} & {{\cal O}\left( {r^{ - 1} } \right)} & {{\cal O}\left( {r^{ - 1} } \right)}  \\
		{} & {} & { {\cal O}\left( {r^{ - 1} } \right)} & {{\cal O}\left( {r^{ - 2} } \right)}  \\
		{} & {} & {} & {{\cal O}\left( r^{-3} \right)}  \\
\end{array}} \right) \,,
\ee 
and for the Maxwell vector field
\be 
a_\mu  {\rm{d}}x^\mu   \sim {\cal O}\left( r \right){\rm{d}}t + {\cal O}\left( {r^{ - 1} } \right){\rm{d}}r + {\cal O}\left( 1 \right){\rm{d}}x + {\cal O} \left(r^{-2}\right) {\rm{d}}\phi \,,
\ee 
so the formalism examined in \cite{Hartman:2008pb,Compere:2012jk} can apply. Accordingly, the most general diffeomorphisms preserving the boundary condition for the spacetime fluctutation can be written as
\be 
K_\varepsilon   = \varepsilon \left( \phi  \right)\partial _\phi   - r\frac{{d\varepsilon \left( \phi  \right)}}{{d\phi }}\partial _r  + {\rm{subleading term}}
\ee 

Now, from the near horizon geometry we have in (\ref{nhmetric}), the associated central charge can be computed using the general formula \cite{Hartman:2008pb}
\be \label{central.charge.gen}
c = c_{grav}  + c_{gauge} \,,
\ee
where
\be \label{c.grav}
c_{grav}  = 3k\int\limits_{ - 1}^{ + 1} {dx\sqrt {\Gamma \left( x \right)\alpha \left( x \right)\gamma \left( x \right)} } \,,
\ee 
and
\be 
c_{gauge} = 0.
\ee 
Inserting the metric component (\ref{nhmetric}) into eq. (\ref{c.grav}) gives us the result
\be\label{central.charge}
c =  \frac{12 m^2 \left\{{a}^{4}-4{b}^{4}\left( 5{l}^{6}{a}^{2}+7{l}^{4}{a}^{4}+3{l}^{2}{a}^{6}+{l}^{8}+4{a}^{8} \right) \right\}}{a^3}\,.
\ee 
As we expected, setting $b=B/2$ and $l=0$ to the last equation gives us the central charge used in the magnetized Kerr/CFT correspondence investigated in \cite{Siahaan:2015xia,Astorino:2015naa}. 

Recall that in section \ref{sec.aspects} we have obtained the Hawking tempereture for a MKTN black hole, which happens to be the same to the temperature of a KTN black hole. In extremal state, this Hawking temperature vanishes, as occurred for the another extremal black holes. However, one can still find the non-vanishing temperature near the horizon, namely the Frolov-Thorne temperature obtained by using the formula \cite{Compere:2012jk}
\be 
T_\phi   = \mathop {\lim }\limits_{r_ +   \to m} \frac{{T_H }}{{\Omega _J^{{\rm{ext}}}  - \Omega _J }} =  - \left. {\frac{{{{\partial T_H } \mathord{\left/
					{\vphantom {{\partial T_H } {\partial r_ +  }}} \right.
					\kern-\nulldelimiterspace} {\partial r_ +  }}}}{{{{\partial \Omega _J } \mathord{\left/
					{\vphantom {{\partial \Omega _J } {\partial r_ +  }}} \right.
					\kern-\nulldelimiterspace} {\partial r_ +  }}}}} \right|_{r_ +   = m} \,.
\ee 
It turns out the Frolov-Thorne temperature for our MKTN case is 
\be\label{FTtemp}
T_\phi   = \frac{a^5}{2\pi \sqrt{a^2 - l^2}  \left\{{a}^{4}-4{b}^{4} \left( 5{l}^{6}{a}^{2}+7{l}^{4}{a}^{4}+3{l}^{2}{a}^{6}+{l}^{8}+4{a}^{8} \right) \right\}} = \frac{1}{2\pi k}\,,
\ee 
where the constant $k$ is given in (\ref{k}). 

Now, having obtained the central charged associated to the symmetry of near horizon of extremal MKTN black hole and the corresponding Frolov-Thorne temperature, we can proceed to get the entropy using the Cardy formula which reads
\be\label{Cardy}
S_{\rm Cardy} = \frac{\pi^2}{3} c T\,.
\ee 
After plugging the central charge $c$ in (\ref{central.charge}) and temperature (\ref{FTtemp}) into (\ref{Cardy}), 
we recover the entropy of extremal MKTN black hole
\be 
S_{\rm ext.} = \frac{{\cal A}_{\rm ext.}}{4} = 2\pi a^2\,.
\ee 
The last equation is just the extremal limit of the entropy given in (\ref{entropy.gen}), i.e. after setting $r_+ = m$ and $m^2+l^2=a^2$. Therefore, here we have managed to reproduce the entropy of an extremal MKTN black hole by using microscopic formula (\ref{Cardy}) according to the Kerr/CFT correspondence \cite{Guica:2008mu}.

\section{Conclusion and discussion}
\label{sec.conclusion}

In this paper we have presented the magnetized version of the KTN spacetime solution of the Einstein-Maxwell theory. The solution is obtained by employing the Ernst magnetization introduced in \cite{Ernst}, which has been used to magnetized some known solutions of the Einstein-Maxwell equations. The novel solution presented in this paper can be considered as a Taub-NUT extension of the previous solution known in literature as the magnetized Kerr spacetime \cite{Ernst}. Some aspects of the spacetime are discussed, namely the area of black hole, squared of Riemann tensor on equator, and the corresponding Hawking temperature.

In addition to the studies on some aspects of the new solution, we also discussed the Kerr/CFT holography associated to the spacetime. The agreement of near horizon metric (\ref{nhmetric}) to the general form discussed in \cite{Hartman:2008pb,Compere:2012jk} allows us to employ the method developed in that works to obtain the corresponding central charge and Frolov-Thorne temperature. In turn, these quantities give us the extremal black hole entropy after using the Cardy formula, which can be viewed as the microscopic calculation for the black hole entropy. This result is a generalization of some previous works on the magnetized Kerr/CFT correspondence \cite{Siahaan:2015xia,Astorino:2015naa}, where now the spacetime is equipped with the NUT parameter.

There are several future projects that can be pursued based on the results presented in this paper. First is to extend the solution obtained here to the charged one, namely the magnetized Kerr-Newman-Taub-NUT solution \cite{SiahaanGhezelbash}. The method would be similar, but the incorporating functions and the results would be more complicated compared to those appearing in this paper. Also discussing the associated Meissner effect as that investigated in \cite{Bicak:2015lxa} for an extremal magnetized black holes with NUT parameter would be interesting.

\appendix

\section{Ernst magnetization}\label{app.ErnstMag}

In this appendix, let us review briefly the Ernst magnetization procedure employed in section \ref{sec.MKerrTN}. The seed metric should be expressible in the form of Lewis-Papapetrou-Weyl (LPW) line element, namely
\be\label{LPWmetric1} 
ds^2  = f\left( {d\phi  - \omega dt} \right)^2 - f^{ - 1} \left( {\rho ^2 dt^2  - e^{2\gamma } d\zeta d\zeta ^* } \right) \,.
\ee
Above, $f$, $\gamma$, and $\omega$ are functions of $\zeta$. Moreover, we have used $-+++$ sign convention for the spacetime, and $^*$ notation represents the complex conjugation. Together with the vector solution $A_\mu dx^\mu = A_t dt + A_\phi d\phi$, the line element above construct the Ernst potentials
\be \label{Ernst.potential.EM}
\Phi  = A_\phi   + i\tilde A_\phi  \,,
\ee 
and
\be \label{Ernst.potential.Grav}
{\cal E} = f + \left| {\Phi } \right|^2   - i\Psi \,.
\ee 
The first potential above is known as the electromagnetic one, and the latter one is gravitational potential. Particularly for the electromagnetic potential (\ref{Ernst.potential.EM}), the imaginary part is obtained by solving
\be \label{eqA}
\nabla A_t +\omega \nabla A_\phi =   - i\frac{\rho }{f}\nabla \tilde A_\phi  \,.
\ee 
On the other hand, the twist potential $\Psi$ in (\ref{Ernst.potential.Grav}) obeys the differential equations
\be \label{eq.Psi}
\nabla \Psi  = \frac{{i f^2 }}{\rho }\nabla \omega + 2i\Phi ^* \nabla \Phi  \,.
\ee 

In terms of Ernst potentials (\ref{Ernst.potential.Grav}) and (\ref{Ernst.potential.EM}), the Einstein-Maxwell field equations can be written as 
\be \label{eq.Ernst.grav}
\left( {{\mathop{\rm Re}\nolimits} \left\{ {\cal E} \right\} + {{\left| \Phi  \right|}^2}} \right)\nabla {\cal E} = \left( {\nabla {\cal E} + 2{\Phi ^*}\nabla \Phi } \right) \cdot \nabla {\cal E}\,,
\ee 
and
\be \label{eq.Ernst.EM}
\left( {{\mathop{\rm Re}\nolimits} \left\{ {\cal E} \right\} + {{\left| \Phi  \right|}^2}} \right)\nabla {\Phi} = \left( {\nabla {\cal E} + 2{\Phi ^*}\nabla \Phi } \right) \cdot \nabla {\Phi}\,,
\ee 
which are known as the Ernst equations. The last two equations are invariant under a type of Harrison transformation
\be \label{magnetization}
{\cal E} \to {\cal E}' = \Lambda ^{ - 1} {\cal E}~~~{\rm and}~~~\Phi  \to \Phi ' = \Lambda ^{ - 1} \left( {\Phi  - b {\cal E}} \right)\,,
\ee
where 
\be \label{LambdaDEF}
\Lambda  = 1 - 2b\Phi  + b^2 {\cal E}\,.
\ee 
The constant $b$ in the equation above represent the strength of external magnetic field where the black hole is embedded. 

Furthermore, the magnetized line element (\ref{LPWmetric1}) resulting from the magnetization (\ref{magnetization}) will contain
\be \label{fp}
f' = {\mathop{\rm Re}\nolimits} \left\{ {{\cal E}'} \right\} - \left| {\Phi '} \right|^{2}  = \left| \Lambda  \right|^{-2} f\,,
\ee 
and
\be \label{wp}
\nabla \omega ' = \left| \Lambda  \right|^2 \nabla \omega  - \frac{\rho }{f}\left( {\Lambda ^* \nabla \Lambda  - \Lambda \nabla \Lambda ^* } \right)\,,
\ee 
while the function $\gamma$ remains unaffected. A little bit more detail, since we would like to express the metric in a Boyer-Lindquist type, the term $d\zeta d\zeta ^*$ in (\ref{LPWmetric1}) can be written as
\be \label{metric2rx}
d\zeta d\zeta ^*  = \frac{{dr^2 }}{{\Delta _r }} + \frac{{dx^2 }}{{\Delta _x }}\,,
\ee 
where $\Delta _r = \Delta _r \left(r\right)$ and $\Delta _x = \Delta _x \left(x\right)$. Note that the corresponding operator $\nabla$ would take the form $\nabla  = \sqrt {\Delta _r } \partial _r  + i\sqrt {\Delta _x } \partial _x $, and we would have $\rho^2 = \Delta_r\Delta_x$. Consequently, (\ref{eqA}) gives us
\be \label{drAt}
\partial _r A_t  =  - \omega \partial _r A_\phi   + \frac{{\Delta _x }}{f}\partial _x \tilde A_\phi  \,,
\ee 
and
\be \label{dxAt}
\partial _x A_t  =  - \omega \partial _x A_\phi   - \frac{{\Delta _r }}{f}\partial _r \tilde A_\phi  \,.
\ee 
The last two equations are useful to obtain the $A_t$ component associated to the magnetized spacetime according to (\ref{magnetization}). To end this appendix, another equations which are important to complete the metric solution are
\be \label{drwp}
\partial _r \omega ' = \left| \Lambda  \right|^2 \partial _r \omega  + \frac{{\Delta _x }}{f}{\mathop{\rm Im}\nolimits} \left\{ {\Lambda ^* \partial _x \Lambda  - \Lambda \partial _x \Lambda ^* } \right\} \,,
\ee 
and
\be \label{dxwp}
\partial _x \omega ' = \left| \Lambda  \right|^2 \partial _x \omega  - \frac{{\Delta _r }}{f}{\mathop{\rm Im}\nolimits} \left\{ {\Lambda ^* \partial _r \Lambda  - \Lambda \partial _r \Lambda ^* } \right\}\,.
\ee 

\section{Components of electromagnetic potential}\label{app.vec}

Here we present the component of electromagnetic Ernst potential $\Phi '  = A_\phi + i {\tilde A}_\phi$ in (\ref{mag.Phi}). The real part reads
\be \label{Ap}
{A_\phi } = {b~{\sum\limits_{j = 0}^6 {{a_j}{x^j}} } \over {\sum\limits_{k = 0}^4 {{b_k}{x^k}} }}
\ee 
where the function $a_j$'s are
\[
a_6 = -a^6 b^2 \Delta_r^2 \,,
\]
\[
a_5 = 6 a^5 b^3 l \Delta_r^2 \,,
\]
\[
a_4 = a^4 \left\{ -{r}^{6}{b}^{2}-15{b}^{2}{r}^{4}{l}^{2} -4{b}^{2}m \left( 3{a}^{2}-10{l}^{2} \right) r^3 +\left(24{b}^{2}{a}^{2}{m}^{2}-36{b}^{2}{l}^{2}{m}^{2}+9{b}^{2}{l}^{4}\right. \right. 
\]
\[
\left. -36{b}^{2}{a}^{2}{l}^{2}+3{b}^{2}{a}^{4}+{a}^{2}
\right) r^2 + -2m \left(12{b}^{2}{l}^{4}+{a}^{2}-42{b}^{2}{a}^{2}{l}^{2}+6{b}^{2}{a}^{4} \right) r
\]
\[
\left. -4\,{b}^{2} \left( {l}^{2}+{a}^{2} \right) ^{2} {m}^{2}+28{b}^{2}{a}^{2}{l}^{4}-9{b}^{2}{l}^{6}+2{a}^{6}{b}^{2}+{a}^{4}-21{b}^{2}{a}^{4}{l}^{2}-{l}^{2}{a}^{2} \right\}\,,
\]
\[
a_3 = 4a^3 l \left\{ 3{b}^{2}{a}^{2}{r}^{4}-12{b}^{2}{a}^{2}m{r}^{3} \left(6{b}^{2}{a}^{4}-2{b}^{2}{a}^{2}{l}^{2}+22{b}^{2}{a}^{2}{m}^{2}+{a}^{2}+2{b}^{2}{l}^{2}{m}^{2} \right) {r}^{2}
\right.
\]
\[
\left. -2m \left( 10{b}^{2}{a}^{4}-12{b}^{2}{a}^{2}{l}^{2}+{a}^{2}-2{b}^{2}{l}^{4} \right) r -2\,{b}^{2} \left( {l}^{2}+{a}^{2} \right) ^{2} {m}^{2}-{l}^{2}{a}^{2}+{a}^{4}-10{b}^{2}{a}^{4}{l}^{2}+7{b}^{2}{a}^{2}{l}^{4}+3{a}^{6}{b}^{2} \right\}\,,
\]
\[
a_2 = a^2 \left\{ 2a^2 b^2 r^6 + {a}^{2} \left( 6{l}^{2}{b}^{2}+1+3{b}^{2}{a}^{2} \right) r^4 - 4{b}^{2}m \left( 2{l}^{4}-3{a}^{4}+2{l}^{2}{a}^{2} \right) r^3 +\left(6{l}^{2}{a}^{2}+24{b}^{2}{m}^{2}{l}^{4}\right. \right.
\]
\[
\left. +18{b}^{2}{a}^{4}{l}^{2}+
48{a}^{2}{b}^{2}{l}^{2}{m}^{2}-36{b}^{2}{a}^{4}{m}^{2}+30{b}^{2}
{a}^{2}{l}^{4}\right) r^2 + 4m \left( 3{a}^{6}{b}^{2}+6{b}^{2}{l}^{6}+{a}^{4}-33{b}^{2}{a}^{4}{l}^{2}-2{l}^{2}{a}^{2} \right)r
\]
\[
\left. -4{b}^{2} \left( 2{a}^{2}+3{l}^{2} \right)  \left( {l}^{2}+{a}^{2} \right) ^{2} m^2-{a}^{2} \left(37{b}^{2}{a}^{2}{l}^{4}-28{b}^{2}{a}^{4}{l}^{2}+{a}^{6}{b}^{2}+3{l}^{4}+{a}^{4}-8{l}^{2}{a}^{2}+6{b}^{2}{l}^{6} \right)
\right\}\,,
\]
\[
a_1 = -2la \left\{ 3{a}^{4}{b}^{2}{r}^{4}+4{a}^{4}{r}^{3}{b}^{2}m + 2{a}^{2} \left( 6{b}^{2}{l}^{2}{m}^{2}+12{b}^{2}{a}^{2}{m}^{2}+3
{b}^{2}{a}^{4}+{a}^{2}+9{b}^{2}{a}^{2}{l}^{2} \right) r^2 \right.
\]
\[ \left. -4m{a}^{2} \left({a}^{2} -9{b}^{2}{a}^{2}{l}^{2}-6{b}^{2}{l}^{4}+3{b}^{2}{a}^{4} \right) r +4{b}^{2} \left( {l}^{2}+{a}^{2} \right) ^{3}{m}^{2}+{a}^{4} \left( 2{a}^{2}-5{b}^{2}{l}^{4}-2{l}^{2}-14{b}^{2}{a}^{2}{l}^{2}+3{b}^{2}{a}^{4} \right)  \right\}\,,
\]
\[
a_0 = -\left\{ a^4 b^2 r^6 +{a}^{4} \left( 1+7{l}^{2}{b}^{2}+2{b}^{2}{a}^{2} \right) r^4 + 4\,{b}^{2}{a}^{2}m \left( {a}^{4}+4\,{l}^{2}{a}^{2}+2\,{l}^{4}
\right) r^3 +\left(16{a}^{4}{b}^{2}{m}^{2}{l}^{2}+{a}^{8}{b}^{2}\right. \right.
\]
\[
\left. +16{a}^{2}{b}^{2}{m}^{2}{l}^{4}+7{a}^{4}{b}^{2}{l}^{4}+{a}^{6}+2{l}^{2}{a}^{4}+4{m}^{2}{l}^{6}{b}^{2}\right) r^2 +2m{a}^{2} \left( 2{a}^{6}{b}^{2}+4{b}^{2}{a}^{2}{l}^{4}+4{b}^{2}{l}^{6}-2{b}^{2}{a}^{4}{l}^{2}+{a}^{4} \right) r
\]
\[
\left. 4{b}^{2} \left( {l}^{2}+{a}^{2} \right) ^{4}{m}^{2}+{l}^{2}{a}^{4}
\left( 3{a}^{2}+{l}^{2} \right)  \left(3{b}^{2}{a}^{2}+{l}^{2}{b}^{2}+1 \right)  \right\}\,,
\]
and the corresponding $b_k$'s are
\[
b_6 = b^4 a^6 \Delta_r^2\,,
\]
\[
b_5 = 6 l a^5 b^4 \Delta_r^2 \,,
\]
\[
b_4 = -b^2 a^4 \left\{ 2{a}^{6}{b}^{2}+2{a}^{4}+84r{b}^{2}m{l}^{2}{a}^{2}+40{b}^{2}m{r}^{3}{l}^{2} -24{b}^{2}mr{l}^{4}-36{m}^{2}{l}^{2}{r}^{2}{b}^{2}-15
{r}^{4}{b}^{2}{l}^{2}\right.
\]
\[
+9{l}^{4}{b}^{2}{r}^{2}-4{b}^{2}{m}^{2}{l}^
{4}-12{b}^{2}{a}^{4}mr-4mr{a}^{2}-36{b}^{2}{a}^{2}{r}^{2}{l}^{2}
+24{b}^{2}{a}^{2}{m}^{2}{r}^{2}-8{a}^{2}{b}^{2}{l}^{2}{m}^{2}-12
{b}^{2}{a}^{2}m{r}^{3}
\]
\[
\left. -4{b}^{2}{a}^{4}{m}^{2}-21{b}^{2}{a}^{4}{l}^{2}+3{b}^{2}{a}^{4}{r}^{2}+2{r}^{2}{a}^{2}-2{l}^{2}{a}^{2}+28{b}^{2}{a}^{2}{l}^{4}-9{l}^{6}{b}^{2}-{b}^{2}{r}^{6} \right\}\,,
\]
\[
b_3 = 4b^2 la^3\left\{ 20{b}^{2}{a}^{4}mr+4mr{a}^{2}+12{b}^{2}{a}^{2}m{r}^{3}-24r{b}^{2}m{l}^{2}{a}^{2}-22{b}^{2}{a}^{2}{m}^{2}{r}^{2} -4{b}^{2}mr{l}^{4} \right.
\]
\[
-2{m}^{2}{l}^{2}{r}^{2}{b}^{2}-3{b}^{2}{a}^{2}{r}^{4}+2{b}^{2}{a}^{2}{r}^{2}{l}^{2}+4{a}^{2}{b}^{2}{l}^{2}{m}^{2}+2{b}^{2}{m}^{2}{l}^{4}-6{b}^{2}{a}^{4}{r}^{2}+2{b}^{2}{a}^{4}{m}^{2}
\]
\[
\left. -2{r}^{2}{a}^{2}-3{a}^{6}{b}^{2}-2{a}^{4}+10{b}^{2}{a}^{4}{l}^{2}+2{l}^{2}{a}^{2}-7{b}^{2}{a}^{2}{l}^{4} \right\}\,,
\]
\[
b_2 = a^2 \left\{ 6{b}^{2}{a}^{2}{l}^{4}-2{b}^{2}{a}^{2}{r}^{4}-16{b}^{2}{a}^{4}{l
}^{2}-28{a}^{6}{b}^{4}{l}^{2}+{a}^{8}{b}^{4}-12r{a}^{6}{b}^{4}m+
132r{l}^{2}{b}^{4}{a}^{4}m +16r{b}^{2}m{l}^{2}{a}^{2} \right.
\]
\[
-24{b}^{4}{m
}^{2}{l}^{4}{r}^{2}-6{b}^{4}{r}^{4}{l}^{2}{a}^{2}+36{b}^{4}{a}^{4}
{m}^{2}{r}^{2}-2{b}^{4}{r}^{6}{a}^{2}-3{r}^{4}{a}^{4}{b}^{4}+8{b
}^{4}{r}^{3}{l}^{2}{a}^{2}m+2{a}^{6}{b}^{2}-48{a}^{2}{b}^{4}{m}^{2
}{l}^{2}{r}^{2}
\]
\[
-12{b}^{2}{a}^{2}{r}^{2}{l}^{2}-8{b}^{2}{a}^{4}mr-
12{r}^{3}{b}^{4}{a}^{4}m+{a}^{4}+12{b}^{4}{m}^{2}{l}^{6}+32{b}^{4}{m}^{2}{l}^{4}{a}^{2}+28{b}^{4}{m}^{2}{a}^{4}{l}^{2}
\]
\[
\left. +8{a}^{6}{b}
^{4}{m}^{2}-24{l}^{6}{b}^{4}mr+8{l}^{4}{b}^{4}m{r}^{3}-30{b}^{4}
{a}^{2}{r}^{2}{l}^{4}-18{b}^{4}{a}^{4}{r}^{2}{l}^{2}+6{b}^{4}{l}^{
	6}{a}^{2}+37{b}^{4}{a}^{4}{l}^{4} \right\}\,,
\]
\[
b_1 = 2la \left\{ {a}^{4}+4{b}^{4}{m}^{2}{l}^{6}-12r{a}^{6}{b}^{4}m-8{b}^{2}{a}^{4}mr+24{b}^{4}{a}^{4}{m}^{2}{r}^{2}+36r{l}^{2}{b}^{4}{a}^{4}m+4{r}^{3}{b}^{4}{a}^{4}m+24{b}^{4}{l}^{4}{a}^{2}mr \right.
\]
\[
+12{a}^{2}{b}^{4}{m}^{2}{l}^{2}{r}^{2}+3{r}^{4}{a}^{4}{b}^{4}+18{b}^{4}{a}^{4}{r}^{2}{l}^{2}+12{b}^{4}{m}^{2}{a}^{4}{l}^{2}+12{b}^{4}{m}^{2}{l}^{4}{a}^{2}+6{r}^{2}{b}^{4}{a}^{6}+4{a}^{6}{b}^{4}{m}^{2}
\]
\[
\left. +4{b}^{2}{a}^{4}{r}^{2}+3{a}^{8}{b}^{4}+4{a}^{6}{b}^{2}-14{a}^{6}{b}^{4}{l}^{2}-4{b}^{2}{a}^{4}{l}^{2}-5{b}^{4}{a}^{4}{l}^{4} \right\}\,,
\]
\[
b_0 = 4{b}^{4}{m}^{2}{l}^{8}+{b}^{4} \left( 8mr{a}^{2}+{a}^{4}+4{m}^{2
}{r}^{2}+16{a}^{2}{m}^{2} \right) {l}^{6}+{b}^{2}{a}^{2} \left( 7{
	r}^{2}{a}^{2}{b}^{2}+8{b}^{2}m{r}^{3} +2{a}^{2}\right.
\]
\[
\left. +8{b}^{2}{a}^{2}mr
+6{b}^{2}{a}^{4}+16{b}^{2}{m}^{2}{r}^{2}+24{b}^{2}{a}^{2}{m}^{2}
\right) {l}^{4} + {a}^{4} \left( 16{b}^{4}m{r}^{3}+9{b}^{4}{a}^{4}+
6{b}^{2}{a}^{2}+4{b}^{2}{r}^{2}+7{b}^{4}{r}^{4}\right.
\]
\[
\left. +16{b}^{4}{m}
^{2}{r}^{2}-4r{b}^{4}{a}^{2}m+16{b}^{4}{m}^{2}{a}^{2} +1 \right) {l}^
{2}+{a}^{4} \left( r+{b}^{2}{a}^{2}r+{b}^{2}{r}^{3}+2{b}^{2}{a}^{2}m \right) ^{2}\,.
\]
The imaginary one can be written as
\be \label{Aptilde}
{{\tilde A}_\phi } = {{-2a^2 b}~{\sum\limits_{j = 0}^3 {{{\tilde a}_j}{x^j}} } \over {\sum\limits_{k = 0}^6 {{{\tilde b}_k}{x^k}} }}\,,
\ee 
where we have ${\tilde a}_j$'s as the followings
\[
{\tilde a}_3 = {a}^{3} \left( m{a}^{2} -m{r}^{2}+m{l}^{2}-2 r{l}^{2} \right) \,,
\]
\[
{\tilde a}_2 = {a}^{2}l \left( {r}^{3} -3m{r}^{2}+2m{l}^{2}-3r{l}^{2}+2m{a}^{2}+3r{a}^{2} \right)\,,
\]
\[
{\tilde a}_1= a \left( 3\,{a}^{2}m{r}^{2}+{a}^{4}m+3\,{a}^{2}m{l}^{2}+2m{l}^{4}+6
{a}^{2}r{l}^{2} \right)\,,
\]
\[
{\tilde a}_0 = l \left( {a}^{2}r{l}^{2}+{a}^{2}{r}^{3}+{a}^{4}m+2{a}^{2}m{l}^{2}+m{l}^{4}+2{a}^{2}m{r}^{2}+{l}^{2}m{r}^{2}-{a}^{4}r \right) \,,
\]
and the associated ${\tilde b}_k$'s read
\[
{\tilde b}_6 = b^4 a^6 \Delta_r^2\,,
\]
\[
{\tilde b}_5 = 6la^5 b^4 \Delta_r^2\,,
\]
\[
{\tilde b}_4 = -b^2 a^4 \left\{ 2{a}^{6}{b}^{2}+2{a}^{4}+84r{b}^{2}m{l}^{2}{a}^{2}+40{b}^{2}m{r}^{3}{l}^{2}-24{b}^{2}mr{l}^{4}-36{m}^{2}{l}^{2}{r}^{2}{b}^{2}-15
{r}^{4}{b}^{2}{l}^{2} \right.
\]
\[
+9{l}^{4}{b}^{2}{r}^{2}-4{b}^{2}{m}^{2}{l}^
{4}-12{b}^{2}{a}^{4}mr-4mr{a}^{2}-36{b}^{2}{a}^{2}{r}^{2}{l}^{2}+24{b}^{2}{a}^{2}{m}^{2}{r}^{2}-8{a}^{2}{b}^{2}{l}^{2}{m}^{2}-12{b}^{2}{a}^{2}m{r}^{3}
\]
\[
\left. -4{b}^{2}{a}^{4}{m}^{2}-21{b}^{2}{a}^{4}{l}^{2}+3{b}^{2}{a}^{4}{r}^{2}+2{r}^{2}{a}^{2}-2{l}^{2}{a}^{2}+28{b}^{2}{a}^{2}{l}^{4}-9{l}^{6}{b}^{2}-{b}^{2}{r}^{6} \right\}\,,
\]
\[
{\tilde b}_3 = -4b^2 la^3\left\{ -20{b}^{2}{a}^{4}mr-4mr{a}^{2}-12{b}^{2}{a}^{2}m{r}^{3}+24r{b}
^{2}m{l}^{2}{a}^{2}+22{b}^{2}{a}^{2}{m}^{2}{r}^{2}+4{b}^{2}mr{l}^{4} \right.
\]
\[
+2{m}^{2}{l}^{2}{r}^{2}{b}^{2}+3{b}^{2}{a}^{2}{r}^{4} -2{b}^{2}
{a}^{2}{r}^{2}{l}^{2}-4{a}^{2}{b}^{2}{l}^{2}{m}^{2}-2{b}^{2}{m}^{2}{l}^{4}+6{b}^{2}{a}^{4}{r}^{2}-2{b}^{2}{a}^{4}{m}^{2}
\]
\[
\left. +2{r}^{2}{a}^{2}+3{a}^{6}{b}^{2}+2{a}^{4}-10{b}^{2}{a}^{4}{l}^{2}-2{l}^{2}{a}^{2}+7{b}^{2}{a}^{2}{l}^{4} \right\}\,,
\]
\[
{\tilde b}_2 = a^2 \left\{ 6{b}^{2}{a}^{2}{l}^{4}-2{b}^{2}{a}^{2}{r}^{4}-16{b}^{2}{a}^{4}{l}^{2}-28{a}^{6}{b}^{4}{l}^{2}+{a}^{8}{b}^{4}-12r{a}^{6}{b}^{4}m+
132r{l}^{2}{b}^{4}{a}^{4}m+16r{b}^{2}m{l}^{2}{a}^{2} \right.
\]
\[
-24{b}^{4}{m}^{2}{l}^{4}{r}^{2}-6{b}^{4}{r}^{4}{l}^{2}{a}^{2}+36{b}^{4}{a}^{4}
{m}^{2}{r}^{2}-2{b}^{4}{r}^{6}{a}^{2}-3{r}^{4}{a}^{4}{b}^{4}+8{b}^{4}{r}^{3}{l}^{2}{a}^{2}m+2{a}^{6}{b}^{2}-48{a}^{2}{b}^{4}{m}^{2}{l}^{2}{r}^{2}
\]
\[
-12{b}^{2}{a}^{2}{r}^{2}{l}^{2}-8{b}^{2}{a}^{4}mr-12{r}^{3}{b}^{4}{a}^{4}m+{a}^{4}+12{b}^{4}{m}^{2}{l}^{6}+32{b}^{4}{m}^{2}{l}^{4}{a}^{2}+28{b}^{4}{m}^{2}{a}^{4}{l}^{2}+8{a}^{6}{b}
^{4}{m}^{2}
\]
\[
\left. -24{l}^{6}{b}^{4}mr+8{l}^{4}{b}^{4}m{r}^{3}-30{b}^{4}
{a}^{2}{r}^{2}{l}^{4}-18{b}^{4}{a}^{4}{r}^{2}{l}^{2}+6{b}^{4}{l}^{6}{a}^{2}+37{b}^{4}{a}^{4}{l}^{4} \right\}\,,
\]
\[
{\tilde b}_1 = 2la\left\{ {a}^{4}+4{b}^{4}{m}^{2}{l}^{6}-12r{a}^{6}{b}^{4}m-8{b}^{2}{a}^{4}mr+24{b}^{4}{a}^{4}{m}^{2}{r}^{2}+36r{l}^{2}{b}^{4}{a}^{4}m+4{r}^{3}{b}^{4}{a}^{4}m \right.
\]
\[+24{b}^{4}{l}^{4}{a}^{2}mr
+12{a}^{2}{b}^{4}{m}
^{2}{l}^{2}{r}^{2}+3{r}^{4}{a}^{4}{b}^{4}+18{b}^{4}{a}^{4}{r}^{2}{l}^{2}+12{b}^{4}{m}^{2}{a}^{4}{l}^{2}+12{b}^{4}{m}^{2}{l}^{4}{a}^{2}+6{r}^{2}{b}^{4}{a}^{6}
\]
\[
\left. +4{a}^{6}{b}^{4}{m}^{2}+4{b}^{2}{a}^{4}
{r}^{2}+3{a}^{8}{b}^{4}+4{a}^{6}{b}^{2}-14{a}^{6}{b}^{4}{l}^{2}-
4{b}^{2}{a}^{4}{l}^{2}-5{b}^{4}{a}^{4}{l}^{4} \right\}\,,
\]
\[
{\tilde b}_0 = 4{b}^{4}{m}^{2}{l}^{8}+{b}^{4} \left( 8mr{a}^{2}+{a}^{4}+4{m}^{2
}{r}^{2}+16{a}^{2}{m}^{2} \right) {l}^{6}+{b}^{2}{a}^{2} \left( 7{r}^{2}{a}^{2}{b}^{2}+8{b}^{2}m{r}^{3}+2{a}^{2}\right.
\]
\[
\left. +8{b}^{2}{a}^{2}mr+6{b}^{2}{a}^{4}+16{b}^{2}{m}^{2}{r}^{2}+24{b}^{2}{a}^{2}{m}^{2}
\right) {l}^{4} +{a}^{4} \left( 16{b}^{4}m{r}^{3}+9{b}^{4}{a}^{4}+
6{b}^{2}{a}^{2}+4{b}^{2}{r}^{2}+7{b}^{4}{r}^{4}\right.
\]
\[
\left. +16{b}^{4}{m}
^{2}{r}^{2}-4r{b}^{4}{a}^{2}m+16{b}^{4}{m}^{2}{a}^{2} +1 \right) {l}^
{2}+{a}^{4} \left( r+{b}^{2}{a}^{2}r+{b}^{2}{r}^{3}+2{b}^{2}{a}^{2}m\right) ^{2}\,.
\]

\end{document}